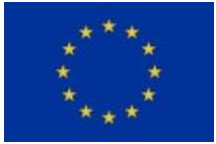
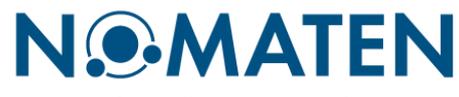

This work was carried out in whole or in part within the framework of the NOMATEN Centre of Excellence, supported from the European Union Horizon 2020 research and innovation program (Grant Agreement No. 857470) and from the European Regional Development Fund via the Foundation for Polish Science International Research Agenda PLUS program (Grant No. MAB PLUS/2018/8), and the Ministry of Science and Higher Education's initiative "Support for the Activities of Centers of Excellence Established in Poland under the Horizon 2020 Program" (agreement no. MEiN/2023/DIR/3795).

The version of record of this article, first published in Materials Science and Engineering: A, Volume 945, November 2025, 149058, is available online at Publisher's website: https://doi.org/10.1016/j.msea.2025.149058

Accepted Version

Publication date: August 2025

European Union, Horizon 2020, Grant Agreement number: 857470 — NOMATEN — H2020-WIDESPREAD-2018-2020, Minister of Science and Higher Education contract no. MEiN/2023/DIR/3795

DOI: https://doi.org/10.1016/j.msea.2025.149058

# Role of chromium oxides and carbides in strengthening CoCrFeNi multi-principle element alloys

Authors: A. Olejarz[1,*], W. Y. Huo[1,2,*], A. Kosińska[1], M. Zieliński[1], T. Stasiak[1], M. Chmielewski[1,3], W. Chmurzyński[1], M. R. Chu[4], M. P. Short[5], Ł. Kurpaska[1]

1 *NOMATEN Centre of Excellence, National Centre for Nuclear Research, A. Sołtana 7, 05-400, Otwock, Swierk, Poland*

2 *College of Mechanical and Electrical Engineering, Nanjing Forestry University, Nanjing, 210037, China*

3 *Lukasiewicz Research Network, Institute of Microelectronics and Photonics, 32/46 Al. Lotników, Warsaw, 02-668, Poland*

4 *Department of Materials Science & Engineering, Massachusetts Institute of Technology, 77 Massachusetts Avenue, Cambridge, MA 02139, USA*

5 *Department of Nuclear Science & Engineering, Massachusetts Institute of Technology, 77 Massachusetts Avenue, Room 24-204 Cambridge, MA 02139, USA*

*Corresponding authors: artur.olejarz@ncbj.gov.pl (A. Olejarz), wenyi.huo@ncbj.gov.pl (W.Y. Huo).

**Abstract:** Multi-principal element alloys (MPEAs) can potentially offer exceptional material properties, but their complex, costly manufacturing limits their scalability. Chemical complexity and complex manufacturing processes lead to the formation of some secondary phases, which have a significant impact on the final properties. In this work, chromium compound dispersoid enhancements (Cr- oxides and carbides) were formed in CoCrFeNi MPEAs to enhance their microstructural and high-temperature mechanical properties. A single FCC phase was observed in the arc melted (AM) samples, chromium oxides were detected in the gas-atomized (GA) samples, and $Cr_2O_3$ with $Cr_{23}C_6$ or $Cr_7C_3$ was found in the mechanically alloyed (MA) samples depending on the sintering temperature. Mechanical tests at room temperature and 575°C, where no phase evolution is expected, showed that the GA samples with oxides achieved enhanced mechanical properties at 575°C. This was co-induced by precipitation strengthening, recrystallization suppression, and twinning-induced plasticity. The MA samples with carbides exhibited high strength but low ductility, with $Cr_7C_3$ outperforming $Cr_{23}C_6$ because of its lower hardness and twinning effects. This work links chromium compound evolution to mechanical performance of MPEAs, offering insights to optimize HEA production for high-temperature applications through controlled phase formation.

**Keywords:** Multi-principal element alloys; CoCrFeNi; arc melting; mechanical alloying; gas atomization; spark plasma sintering

# 1. Introduction

Multi-principal element alloys (MPEAs) have gained attention because of their superior material properties over wide temperature ranges [1–4], corrosion resistance [5,6] or even superb radiation damage resistance [7–9]. The concept of MPEAs involves obtaining a single-phase matrix while mixing several elements, with none constituting a majority of the concentration [3,4]. Recent research has shown that MPEAs can maintain, without any sacrifice, the desired properties originating from individual ingredients [2–4]. The improvements in mechanical properties are limited not only to properties such as increased hardness or tensile strength [10], but also to improved fatigue [11] or enhanced ductility [12].

For the last twenty years, many different chemical compositions have been investigated as MPEAs [2,4,10]. Both FCC [5,13,14] and BCC [15,16] structures have been extensively studied, and MPEAs with HCP structures have also been reported [17], while dual-phase MPEAs are very common due to their superb mechanical properties. FCC-MPEAs may be treated as a next-step class of materials following austenitic stainless steels or nickel-based superalloys, which contain many alloying elements but have a single base chemical element [2]. In turn, FCC-MPEAs are based on a similar chemical composition, but their properties are improved because of a sluggish diffusion coefficient, severe lattice distortion, or solid solution strengthening [2,3,18]. Thus, FCC-ordered MPEAs might be a great substitute, if the chemical composition and manufacturing technique are correctly chosen and demonstrated at scale.

Despite the exceptional properties of BCC-MPEAs over a wide temperature range, the application of refractory elements, such as W, Ta, and Hf, on a mass scale presents some difficulties, including economic risks due to scarcity or negative environmental impact [19]. The use of dual-phase MPEAs results in a great strength-elongation trade-off, whereas other parameters, including temperature stability, are missing [2,20]. On the other hand, rare-earth elements are desirable for forming HCP-MPEAs, which makes their application challenging [21]. Some promising results might be found for Ti-based HCP-MPEAs [22]. However, HCP-Ti alloys are known for their high processing cost in traditional subtractive manufacturing, which limits their applicability in several industrial fields [23].

Notably, most FCC-MPEAs to date are based on the first developed HEA, i.e., the Co-Cr-Fe-Mn-Ni system [3,4]. This equiatomic composition often serves as a benchmark for more complex alloys or different alloy chemistries. Equiatomic CoCrFeNi still attracts significant attention because of its promising properties over a wide temperature range [24–26]. Importantly, this system does not require expensive alloying elements such as Zr, V, or Re [9,19,26,27]. Moreover, the promising oxidation behavior of the CoCrFeNi MPEA allows engineers to produce the alloy via a less stringent process than that of CoCrFeMnNi, where Mn is a highly oxidizing element [5,28]. Recent research has shown that even a slight addition of alloying elements may greatly enhance the mechanical and functional properties of a material [29–31]. In turn, this composition is a great point to start the evaluation of MPEAs.

The equiatomic CoCrFeNi system may therefore constitute a great alternative for austenitic stainless steels or nickel superalloys for applications in extreme environments. However, the choice of manufacturing technique is no less crucial than the chemical composition [32,33]. The use of different manufacturing techniques might even broaden the perspective of potential applications for the same chemical composition.

MPEAs are usually manufactured starting from pure elements in the form of powder or coarse pieces that are commonly consolidated through melting or powder metallurgy techniques [1,4]. Among the melting-based methods, arc melting under protective atmospheres followed by rapid cooling is used to obtain high purity alloy pieces, avoid intermetallic phase formation [34] and limit the uncontrolled growth of oxides [4]. This is because while intermetallics may be thermodynamically stable, their nucleation and growth can be kinetically constrained by limiting time at high temperatures during processing if the application temperature is significantly lower. However, materials formed in this way often present dendritic structures that deteriorate the mechanical properties [35]. This issue may be overcome by subsequent thermomechanical processes which significantly improve structural and microstructural uniformity and, thus, mechanical properties [36,37].

An interesting alternative for the arc melting technique is powder metallurgy (PM). In this method, material in the form of powder must be alloyed before the consolidation process occurs. Among the many different alloying processes, mechanical alloying (MA) and gas atomization (GA) present the broadest spectrum of potential applications [1]. In the first case, MA is used to form alloy powders via the milling of pure, elemental powders in a milling jar [1]. On the other hand, in the GA, pieces of pure elements are melted, followed by squeezing through a nozzle. A protective gas stream then pulverizes the molten drops of the alloy [38].

GA therefore lies between arc melting and PM methods. Although the GA consolidation technique is similar to that of PM, essential alloying takes place in a liquid state. The consolidation process might take place in several different ways; however, spark plasma sintering (SPS) is the most popular method [1,39,40]. It is easily scalable, which drives interest from industrial partners. In this process, the internal heating source formed between compressed powder particles is enhanced by the external heating source by the direct current through the graphite die [39]. During the sintering process, the surface of the particles is melted, which enables their consolidation. Notably, the greatest disadvantage of PM techniques is that many parameters need to be set before the process starts [1,41]. Consequently, there is a risk of the formation of secondary phases that lead to a decrease in the mechanical properties. On the other hand, the main problem in the melting process is that maintaining a high vacuum is difficult, which makes industrial scalability challenging [34].

The above considerations lead to the conclusion that process optimization of each method must be performed carefully. However, the basic difference and limits of each process might be found even if the manufacturing process is not fully optimized. On the other hand, a literature review revealed that comparisons of both techniques using the same chemical composition are extremely rare [42].

The abovementioned factors encouraged us to investigate the structural evolution of a well-studied MPEA as a function of the manufacturing method. CoCrFeNi was chosen for its many studies and excellent properties for wide potential industrial applications. Four different samples were subjected to different manufacturing methodologies. In the first case, the sample was manufactured by arc melting, followed by homogenization, cold rolling and recrystallization processes to obtain a single FCC structure with fine grains. Furthermore, the GA powder was subjected to a short milling process to reduce the spherical shape of the powder particles. The powder was then consolidated via SPS. The last step of manufacturing was homogenization under the same conditions as those in the arc melted case. Finally, pure-elemental powders were mixed in a WC milling jar, followed by MA in a planetary ball mill. The ball-to-powder ratio was reduced to obtain a similar number of samples and improve the productivity of the milling process. MA powder was then synthesized via SPS using the same parameters as those used in the GA+SPS case. Two different annealing temperatures (850°C and 1050°C) were employed to promote different types of precipitates. In this study, a scanning electron microscope (SEM) equipped with Energy-Dispersive X-ray Spectroscopy (EDS) and Electron Backscatter Diffraction (EBSD) detectors, transmission electron microscopy (TEM), and X-ray diffraction (XRD) were used to analyze the structural and microstructural changes in the materials. Moreover, different mechanical and thermal properties have been studied. Mini-tensile tests were performed to obtain the mechanical response of the materials. This work aims to elucidate the advantages and limitations of all methods used for the manufacturing of MPEAs and to determine their impacts on the mechanical properties at room and elevated temperatures.

## 2. Experimental

2.1 Sample preparation

In this work, three different manufacturing paths were employed. First, CoCrFeNi samples were prepared from high-purity elements via an arc melting furnace equipped with a copper mold. The chamber was evacuated to $5 \times 10^{-5}$ mbar and back-filled with argon (purity: 99.99999%) to a pressure of 600 mbar three times. To ensure chemical homogeneity, the samples were melted and flipped at least five times. The samples were homogenized in a muffle furnace under an air atmosphere for 4 hours at 1200 °C to remove the dendritic structure, followed by water quenching. The thickness of the homogenized ingots was reduced by ~70% via cold rolling in several steps (approximately 0.3 mm/step), and the samples were then recrystallized at 850 °C for 30 min. In the following part of the manuscript, this sample is referred to as AM.

The second set of samples was produced from GA powder produced by American Elements (oxygen and nitrogen contents <900 ppm). The maximum size of the atomized powder was 25 µm. The powder was milled for 10 hours to decrease the spherical shape of the particles, which impedes the sintering process. The milling speed was 250 rpm. The ball-to-powder ratio was 5:1. After the short milling process, the samples were consolidated via SPS. The sintering process was performed under a pressure of 50 MPa at 950°C for 10 minutes. The heating rate was 100°C/min. Afterwards, the samples were homogenized via the same parameters as those used for the arc-melted samples to compare both structures clearly. This sample is referred to as the GA.

Mechanically alloyed MPEA powders of CoCrFeNi were obtained in two milling steps. The Cr and Ni powders were milled first to reduce the size of the particles and promote diffusion between the two elements with the lowest diffusion coefficients. A similar methodology used by Suprianto et al. [43] was employed, resulting in uniform mixing of both elements. The milling was performed in a WC jar using Φ7 WC balls for 10 hours with a milling speed of 350 rpm and a BPR of 10:1. Short milling breaks of 5 min were employed after each 15 min of milling to cool the system. A total of 1.5 wt.% N-heptane ($C_7H_{18}$) was added as a process control agent (PCA) to avoid cold welding of powder particles. After this process, Co and Fe were added, and the powder was milled again with the same parameters, except for the milling time, which was prolonged for up to 45 hours, and the BPR, which was reduced to 5:1. The sintering process was again performed via SPS under a pressure of 50 MPa at 950°C for 10 minutes. The heating rate was 100°C/min. The samples produced in this way were then divided into two different sets. Our recent work revealed that different annealing temperatures promote the formation of different carbides [44]. On the basis of this approach, we employed two different annealing temperatures. The first sample was annealed at 850°C for 6 hours (denoted in the manuscript as MA-850), and the second sample was annealed at 1050°C for 6 hours (denoted in the manuscript as MA-1050) in the same muffle furnace as in AM and GA. After consolidation and thermomechanical treatment, the samples were cut and ground with sandpapers #320 to #2500, polished with 3- and 1- µm diamond pastes and then electropolished with a mixture of 8% perchloric acid with 92% ethanol. The electropolishing process was conducted at 25 V for 20 s at 0°C. The samples used for the tensile tests were cut via an AgieCharmilles E350 electric discharge machine to obtain dog-bone-shaped samples (roughness: Ra3) (SSJ-3 type, commonly used when only small quantities of materials could be produced [45,46]).

2.2 X-ray diffraction

Diffraction pattern (DP) acquisition was performed via a divergent X-ray beam and Bragg-Brentano parafocusing geometry on a Bruker D8 Advance diffractometer with a θ/θ goniometer with a 280 mm radius and a Cu sealed X-ray tube ($\lambda_{CuK\alpha1}$ = 1.5406 Å). Data were acquired by the LYNXEYE XE-T detector working at high energy resolution (ΔE < 380 eV at 8 keV, energy window optimized for $CuK_\alpha$ lines) in 1D mode without a Ni filter, and its matrix (192 strips, each 0.075 mm wide) covered 2.941° of 2θ. Diffraction patterns were collected in the 20° - 145° and 30° - 67° 2θ ranges. The latter featured improved counting statistics to facilitate the analysis of the minor carbide and oxide phases. The Bruker DIFFRAC.EVA program with the database of diffraction standards ICDD PDF5+ 2023 [47] and DIFFRAC.TOPAS programs were used for phase identification and refinement of the crystal structure models of the identified phases. The TOPAS program uses the fundamental parameter profile fitting (FPPF) approach to account for instrumental effects and the Rietveld approach [48,49] to optimize the model's crystal structure parameters.

2.3 Microstructural studies

Microstructural studies were performed via a ThermoFisher Helios 5 UX scanning electron microscope (SEM). Microstructural observations were performed on a backscattered electron (BSE) channeling contrast. An EDAX Octane Elite Plus energy dispersive X-ray spectroscopy (EDS) system and an EDAX Velocity Pro electron EBSD camera were employed to study the chemical homogeneity and grain size discrepancies between different manufacturing processes. EBSD mapping was carried out with a step size of 0.1 µm. The obtained data were analyzed via EDAX OIM Analysis 8 software. Points with a confidence index below 0.1 were eliminated from the calculations.

Transmission electron microscopy (TEM) studies were performed using a JEOL JEM F200 instrument operating at 200 kV. The samples for TEM analysis were prepared via the lift-out focused ion beam (FIB) technique. The final thinning of the samples was performed at a 2 keV beam energy to reduce the density of defects introduced by the $Ga^+$ ions.

2.4 Mechanical tests

Tensile tests were carried out on SSJ-3 dog bone-shaped specimens [45,46] via a Zwick/Roell Z020 Allround-Line testing machine at a strain rate of $1 \times 10^{-3}$ $s^{-1}$. The strain during testing was measured via a laser-based digital image correlation (DIC) system. Two different temperatures, 25°C and 575°C, were applied during the tensile tests to study the mechanical behavior at different temperatures. The higher temperature was chosen because of the chemical instabilities of precipitates at higher temperatures [50,51]. This phase evolution may lead to significant changes in the mechanical properties, leading to an inability to directly compare the results. The samples were heated to 575 °C at a rate of 20°C/min and held for 3 minutes to ensure a homogeneous temperature distribution. A thermocouple remained in contact with the tensile specimen throughout the heating process to ensure the desired heating rate and test temperature were achieved. Three samples were tested at each temperature to ensure the repeatability of the obtained results. EBSD maps and SEM fractographic studies were performed on the specimens after tensile tests. Thin sections for TEM observations were also cut from the deformed dog bone samples.

## 3. Results

3.1 Structural investigations

The diffraction patterns of all the samples described above are presented in Fig. 1 (a). For the AM and GA samples, no secondary phases were detected, as shown in Fig. 1 (b) and (c) respectively. The MA-850 and MA-1050 specimens present an FCC structure (~90 wt.%) which is accompanied by chromium carbides and oxides, including $Cr_2O_3$, $Cr_{23}C_6$, and $Cr_7C_3$. Furthermore, depending on the annealing conditions, different types of carbides were promoted: $Cr_{23}C_6$ was found after annealing at 850°C (see Fig. 1 (d)), whereas at 1050°C, only $Cr_7C_3$ was detected (Fig. 1 (e)). Moreover, $Cr_2O_3$ was found for both MA samples. No signal of chromium oxides for the AM and GA samples was recorded, which suggests that the amount of that phase is lower than the detection level (usually below 2%), indicating the amplitude of noise of the baseline. The lattice parameter of the FCC matrices in all samples is, on average, 3.573 (±0.002) Å, which agrees with the widely reported value for this chemical composition [26,52,53]. The FCC phases in all the samples require the Stephens model, which includes crystal lattice strain anisotropy and deviations from the average value in different crystallographic directions, because the FWHMs of consecutive peaks do not smoothly follow the tan(2θ) correlation. The phase compositions of all the manufactured samples obtained via X-ray diffraction (XRD) are shown in Table 1.

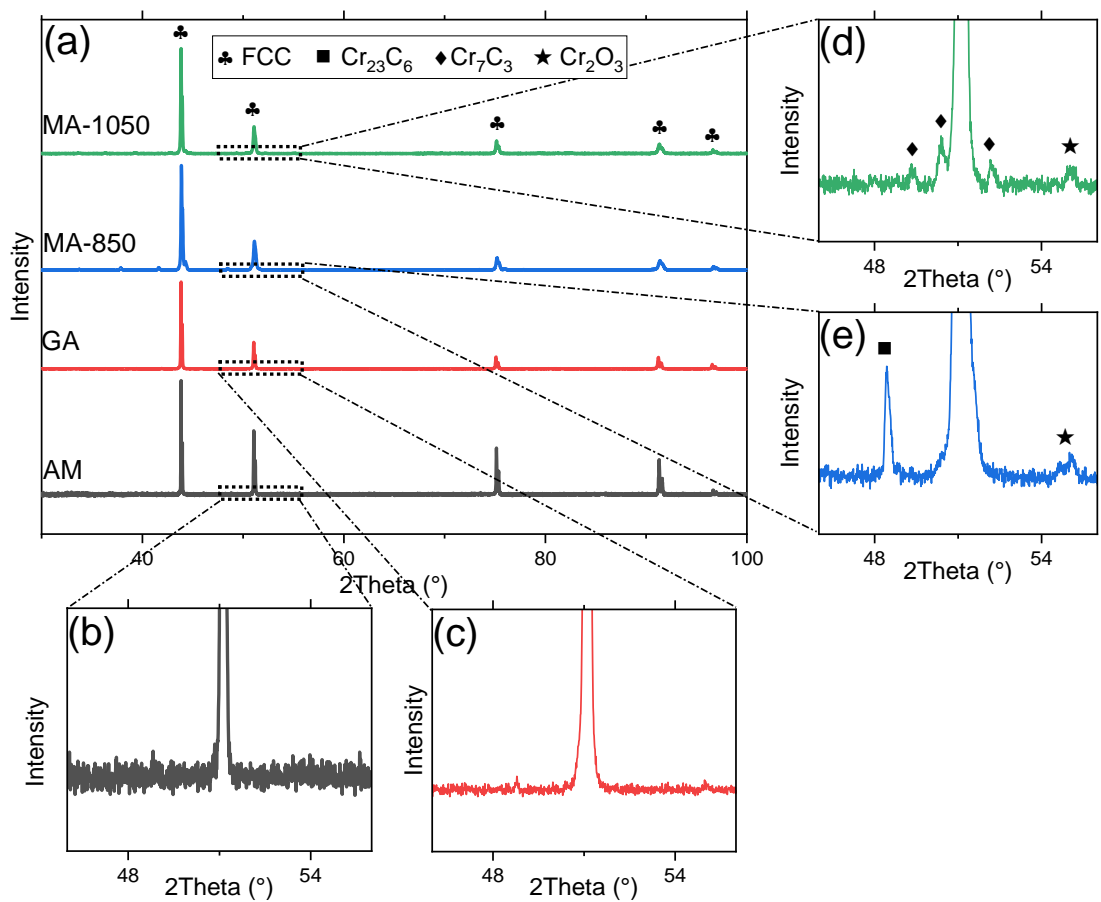

Fig. 1 (a) X-ray diffraction patterns of all synthesized samples with magnification in the specified region (47°-56°) in the AM sample (b), GA sample (c), MA-850 sample (d) and MA-1050 sample (e).

Tab. 1 Secondary phase contents of the different samples from PXRD patterns (* - from the microstructural studies)

|  | Phase composition (wt.%) | | | |
| --- | --- | --- | --- | --- |
|  | FCC | $Cr_2O_3$ | $Cr_{23}C_6$ | $Cr_7C_3$ |
| AM | 100.00 | - | - | - |
| GA | 100.00 | - | - | - |
|  | (98.50*) | (1.50*) | - | - |
| MA-850 | 87.86±0.11 | 1.66±0.08 | 10.48±0.08 | - |
| MA-1050 | 90.50±0.30 | 2.40±0.03 | - | 7.02±0.19 |

## 3.2 Microstructural studies

SEM images of all the manufactured samples are shown in Fig. 2 (a-d). The microstructure of the AM sample presented in Fig. 2 (a) consists of uniform grains. No secondary phases distributed in the matrix were detected. The GA sample presents a similar microstructure, which is visible in Fig. 2 (b). However, some fine oval particles were observed in the matrix phase, clearly suggesting secondary phase formation (black dots). They are located primarily in the vicinity of the grain boundaries, however, a low amount is detected in the matrix. The blue arrows and circles indicate the presence of these particles. To establish the origin of this phase, which not found by XRD, TEM-EDS studies were employed as described in the following section. On the other hand, in the MA samples, different types of secondary phases were observed. As mentioned in the experimental section, BSE contrast was employed. The contrast variation in BSE imaging is determined by the atomic number (Z) of the elements. This Z-contrast is desirable for visualizing the microstructural changes in the frame of chemical composition differences. This means that any change in contrast is related to atomic number, such that heavier elements appear lighter. In this way, it can be found that particles formed in MA samples are composed of lighter elements than the matrix phase. In MA-850 (see Fig. 2 (c)), high contrast, light gray particles can be spotted (marked by red circles), whereas in MA-1050 (Fig. 2 (d)), low contrast, dark gray particles (marked with yellow circles) can be observed. Both images demonstrate that the particles are regionally agglomerated and distributed in a matrix phase. The average grain size of the MA samples is significantly lower than that of the AM and GA samples.

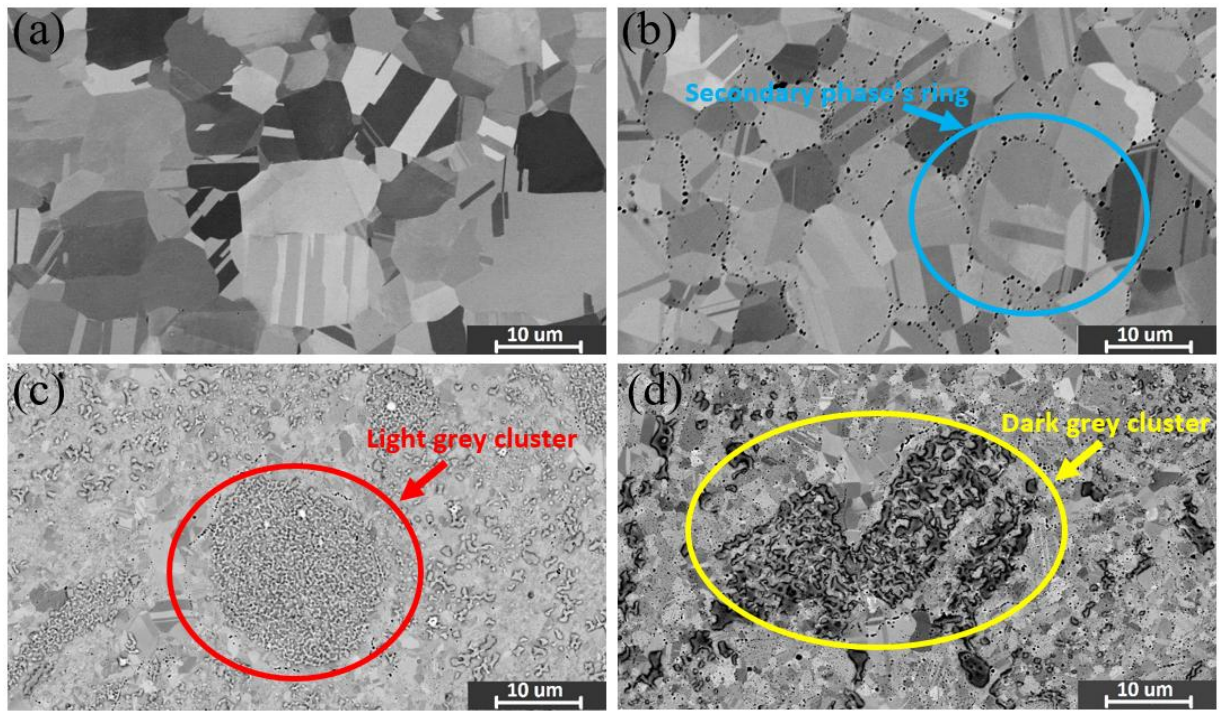

Fig. 2 Microstructures of the samples manufactured via different processing routes: (a) AM, (b) GA, (c) MA-850 and (d) MA-1050.

Furthermore, EBSD was employed to compare the average grain size of all the manufactured samples. The results for all average grain sizes are shown in Tab. 2, whereas the EBSD maps are presented in Fig. 3. Notably, both MA samples exhibit average grain sizes one order of magnitude smaller than those of the AM and GA samples, with an average grain size of ~1 μm. Notably, both mechanically-alloyed samples, especially MA-850, present a high standard deviation of the average grain size. There might be several reasons for this behavior. First and foremost, the MA+SPS process promotes fine-grain particles, with further rapid consolidation, which hinders grain growth. Moreover, the formation of chromium carbides in a matrix inhibits grain growth during the annealing process. Ultimately, the formation of chromium clusters leads to a bimodal grain size distribution. In Fig. 3, the black regions are not identified. They appear to be unanalyzed defects, but the caption does not clarify their nature. The AM and GA samples present similar average grain sizes. Some black spots are visible on the EBSD maps. The size and distribution of these spots correspond well with the SEM images, suggesting that they represent chromium oxides (GA) and chromium carbides (MA-850 and MA-1050). Each sample shows twins that develop preferentially in the large grains.

Tab. 2 Average grain size with standard deviation of all the manufactured samples

| Sample | AM | GA | MA-850 | MA-1050 |
|---|---|---|---|---|
| Grain size [µm] | 10.70±5.31 | 7.88±4.21 | 0.76±0.47 | 1.30±0.79 |

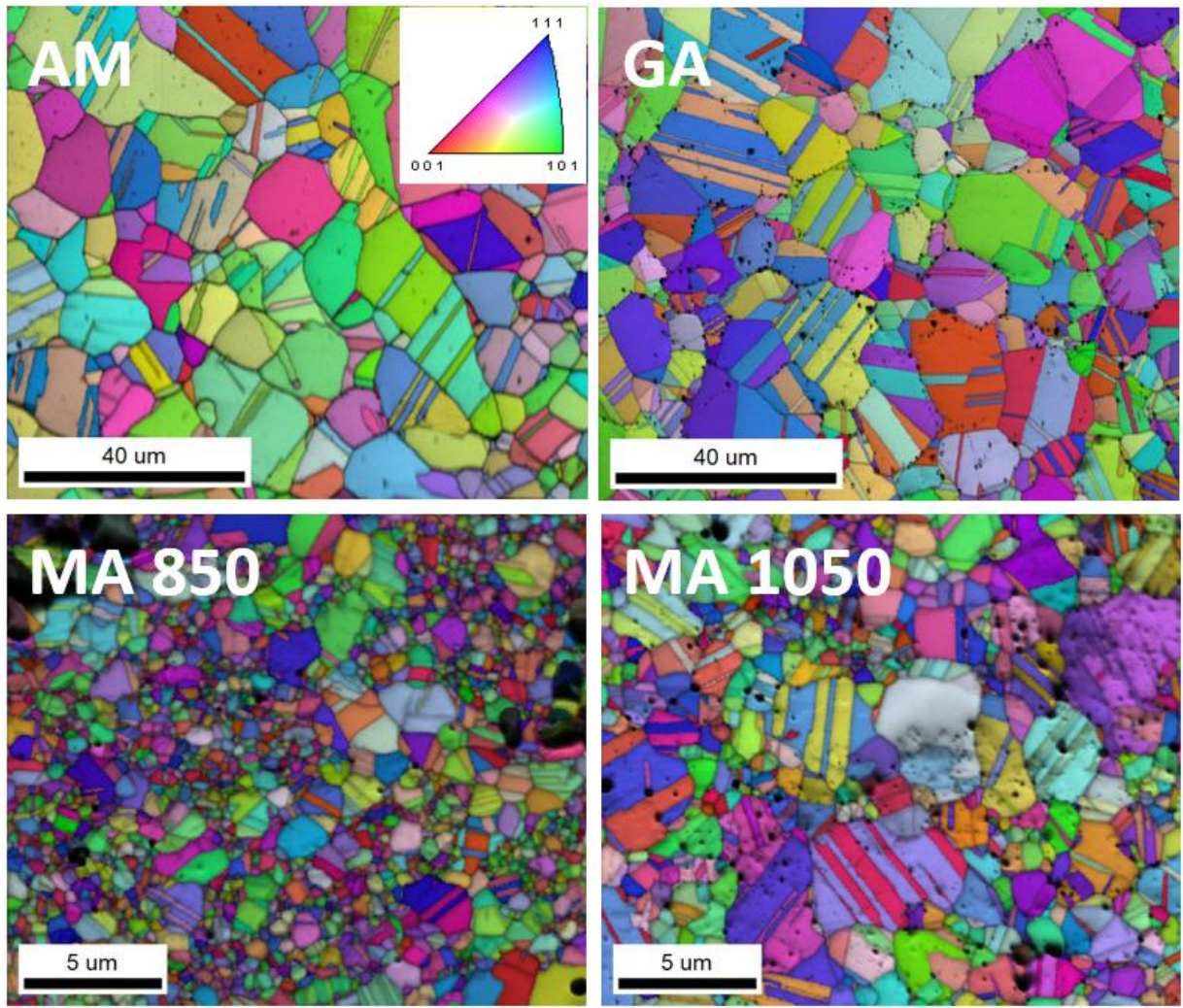

Fig. 3 EBSD inverse pole figure maps of all the manufactured samples

The EDS technique was employed to study the element distribution in the matrix phases for all the manufacturing routes. The average element distribution in the matrix phase determined via point analysis is presented in Fig. 4. At least three points were taken into account. Notably, the distribution of elements in the AM matrix phase is even. Similar observations could be made for the GA sample, which suggests that only a slight amount of alloying elements were consumed during secondary phase formation. On the other hand, visible chromium depletion in the MA samples (MA-850 and MA-1050) can be observed.

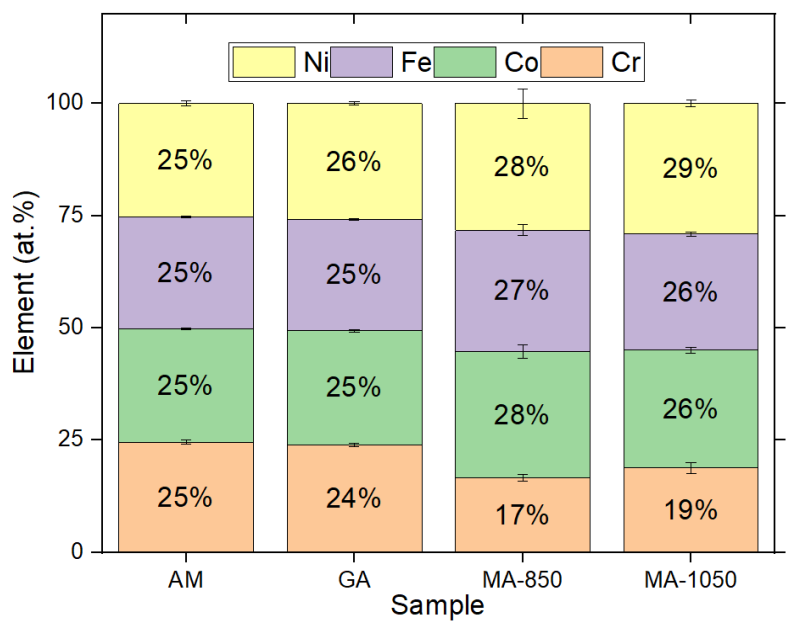

Fig. 4 EDS analysis of the matrix phase with standard deviation in all the manufactured samples

As previously mentioned, some precipitates were found in the GA sample but were not detected via XRD. EDS analysis via TEM as shown in Fig. 5, revealed chromium oxide formation. More detailed analysis via ImageJ software was performed on the SEM image (shown in the Supplemental File), which allowed us to conclude that the chromium oxides constitute approximately 1.5%$_{wt.}$ of the whole sample volume of the GA (see Tab. 1).

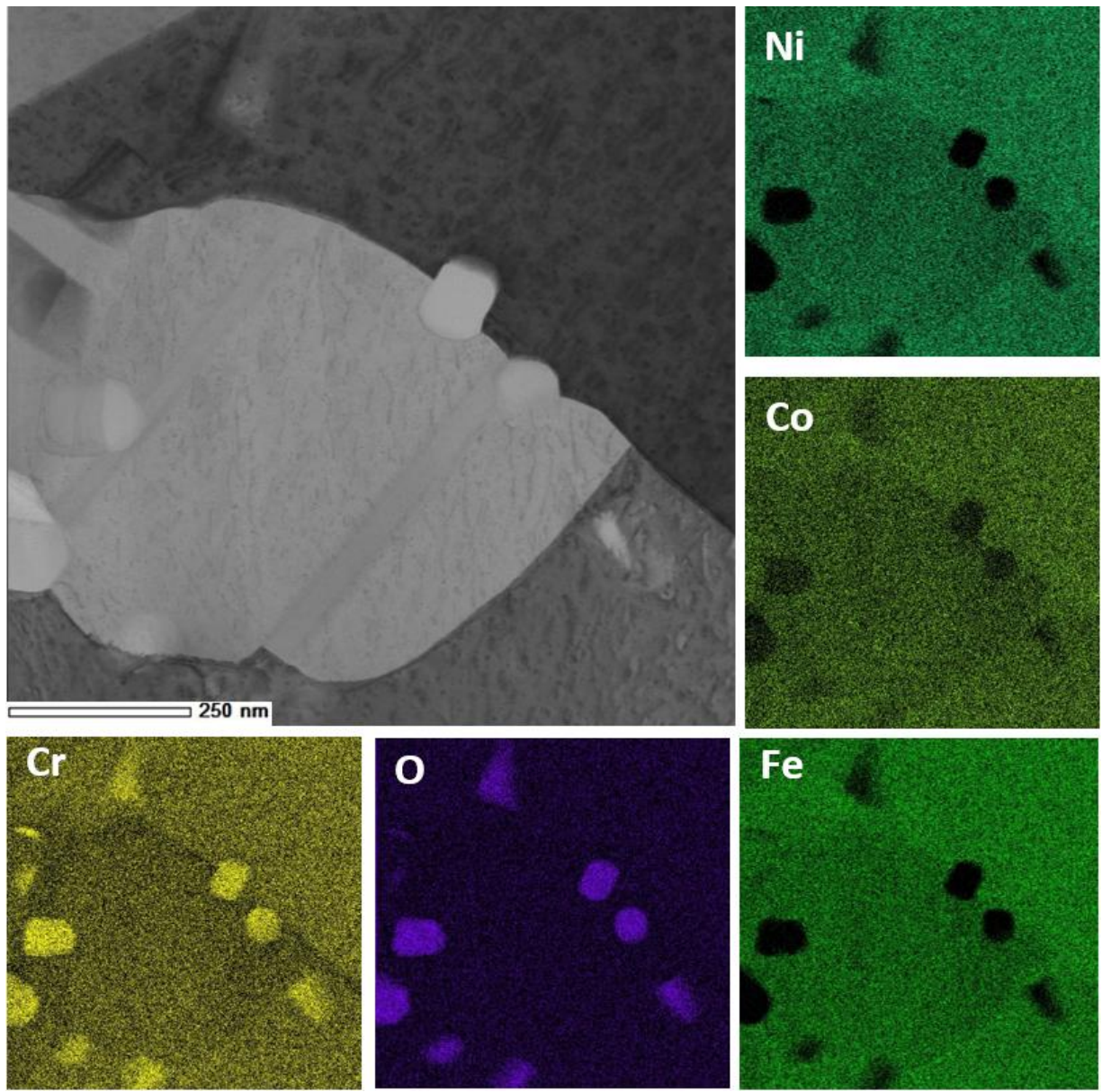

Fig. 5 STEM bright-field image of the GA sample with EDS results showing clear evidence of Cr-oxides at/near the grain boundaries.

3.3 Mechanical properties

Fig. 6 (a) shows the engineering stress-strain curves of all the manufactured samples recorded at room temperature. The ultimate tensile strength (UTS) increases, and the elongation decrease strongly corresponds with the presence of reinforcing phases. The AM sample presents the lowest strength and a large elongation. The GA sample presented a slightly higher yield strength (YS) and UTS and maintained satisfactory elongation. At the same time, both MA samples demonstrate a much greater tensile strength, but their ductility is significantly reduced. Notably, MA-850 fractures almost immediately after reaching the plastic region, whereas MA-1050 exhibits approximately 6% elongation. All the results described above are presented in Tab. 3. To compare the hardening ability of all the manufactured samples, the strain hardening rates were calculated and shown in Fig. 6 (b). The AM and GA samples present a significantly greater strain hardening ability than the MA samples do. Moreover, the MA-850 curve confirms that there is no strain hardening effect before fracture, whereas the MA-1050 curve presents a slight hardening effect.

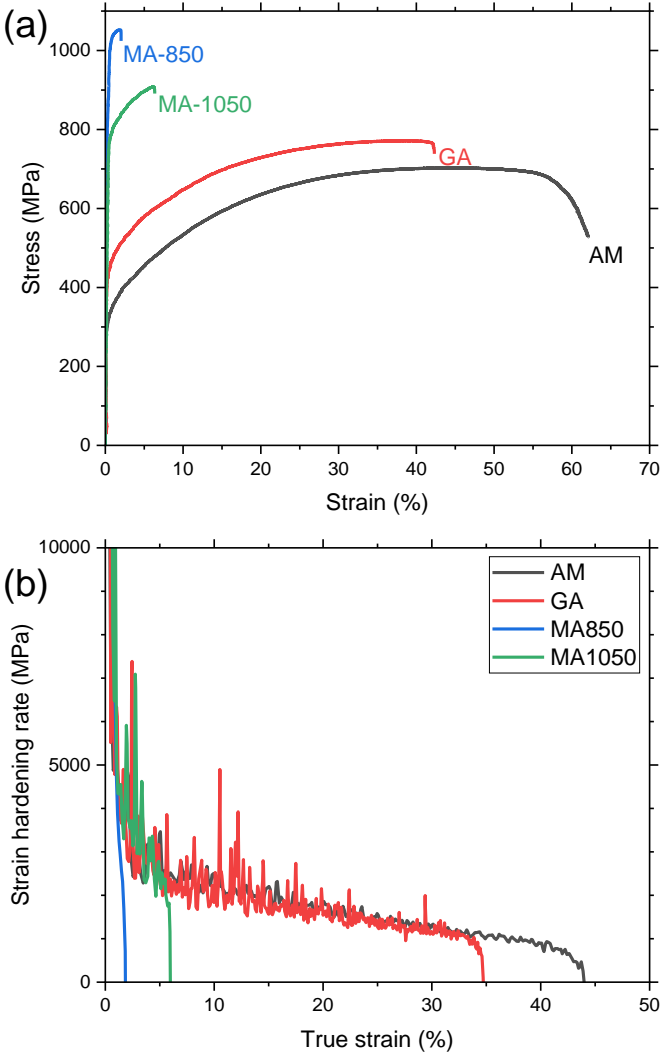

Fig. 6 (a) Representative engineering stress-strain curves of CoCrFeNi samples manufactured via different methods at room temperature. (b) Strain hardening rates of all samples at room temperature.

The engineering stress-strain curves obtained at 575°C are shown in Fig. 7 (a). Although the tensile tests performed at elevated temperatures caused the mechanical properties to decrease, all the materials behaved differently than they did at room temperature. The MA samples demonstrate a decrease in the YS and UTS, but their elongations remain intact. On the other hand, the AM sample has a lower YS and UTS as well as a large decrease in elongation. Moreover, a modest decrease in the UTS of the GA with elongation was recorded. Finally, some local curve fluctuations, commonly known in the literature as serration behavior [54], of AM and GA at 575°C can be observed. However, different types of serrations were detected depending on the manufacturing process. For the AM sample, cycle behavior is visible (Fig. 7 (b)), followed by a completely random way of behaving near the UTS point (Fig. 7 (c)). Moreover, the GA presents a "stairs" behavior, which starts at approximately 17% elongation, as shown in Fig. 7 (d). These effects suggest different deformation mechanisms at elevated temperatures. All the parameters are listed in Tab. 3. The strain hardening rates (Fig. 7 (e)) again confirm the greater hardening ability of AM and GA than MA-850 and MA-1050. One can see several sharp peaks on the GA curves, whose positions correspond strictly with the stairs found in Fig. 7 (d).

Tab. 3 Mechanical properties of all the manufactured MPEAs (each material was tested three times under the given experimental conditions)

| Sample | Temperature | YS [MPa] | UTS [MPa] | ε [%] |
|---|---|---|---|---|
| AM | RT | 363 ± 25 | 693 ± 14 | 56 ± 10 |
| | 575°C | 249 ± 6 | 454 ± 2 | 33 ± 1 |
| **GA** | RT | 430 ± 69 | 757 ± 29 | 42 ± 2 |
| | 575°C | 313 ± 5 | 643 ± 4 | 40 ± 5 |
| **MA-850** | RT | 1037 ± 27 | 1077 ± 29 | 2 ± 1 |
| | 575°C | 801 ± 30 | 809 ± 18 | 2 ± 1 |
| **MA-1050** | RT | 855 ± 65 | 927 ± 31 | 5 ± 1 |
| | 575°C | 662 ± 37 | 738 ± 46 | 7 ± 2 |

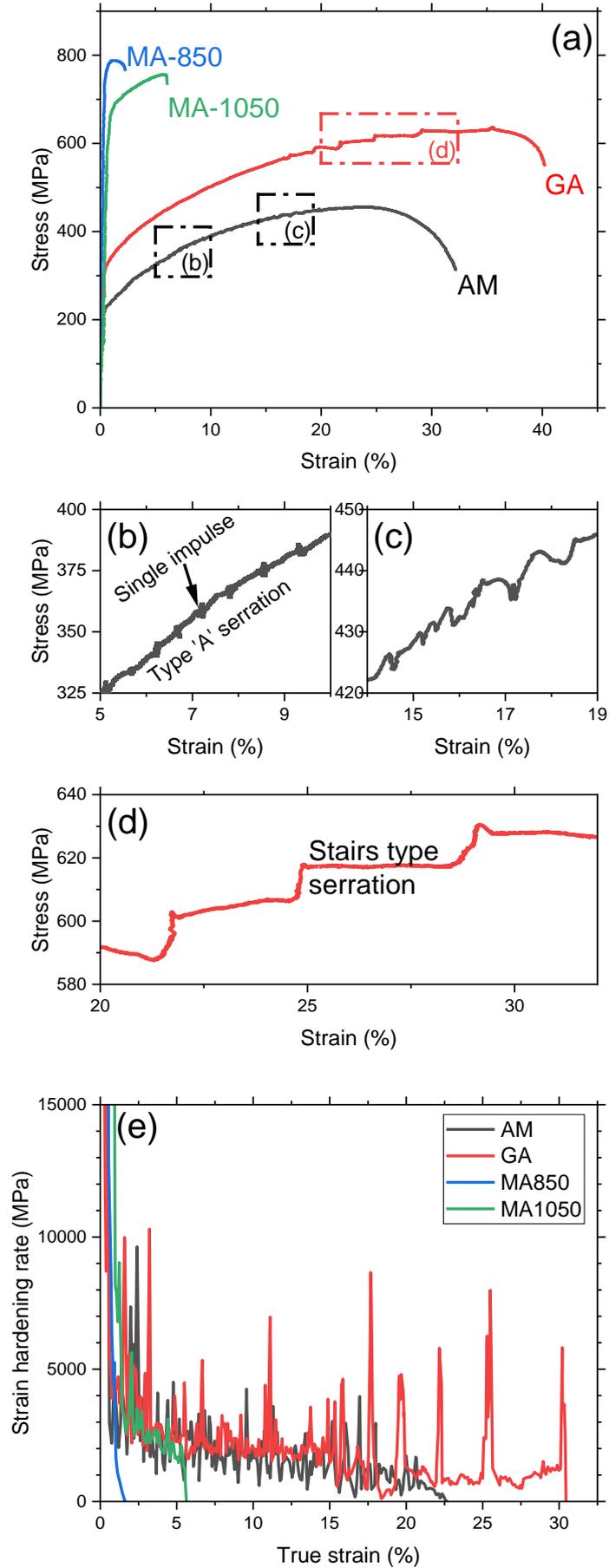

Fig. 7 (a) Engineering stress–strain curves of CoCrFeNi samples manufactured via different methods at 575°C. (b) Highlighted region 5-10% elongation of the AM sample. (c) Highlighted region 14-19% elongation of the AM sample. (d) Highlighted region 20-32% of the GA sample strain. (e) Strain hardening rate of all samples at 575°C.

The fracture surfaces of all the manufactured samples at room temperature (RT) are presented in Fig. 8 (a-d). Depending on the investigated specimen, different deformation mechanisms can be described. The tensile fracture surfaces of AM and GA, presented in Fig. 8 (a) and (b), respectively, exhibit abundant dimples, which is typical for ductile fracture behavior. On the other hand, both the MA-850 and MA-1050 (Fig. 8 (c) and (d), respectively) samples can be characterized by cleavage facets, which are typical of brittle materials. Carbide clusters may act as weak points, as supported by the presence of cracks within these clusters. Moreover, EDS maps were generated to confirm the presence of chromium oxides and carbides in the GA and MA samples, respectively. Based on previous studies, fine Cr-rich precipitates are related to $Cr_2O_3$ in GA, whereas a large Cr cluster in MA samples represents chromium carbides. The detailed EDS maps are attached in a supplemental file (Figs. S3 and S4).

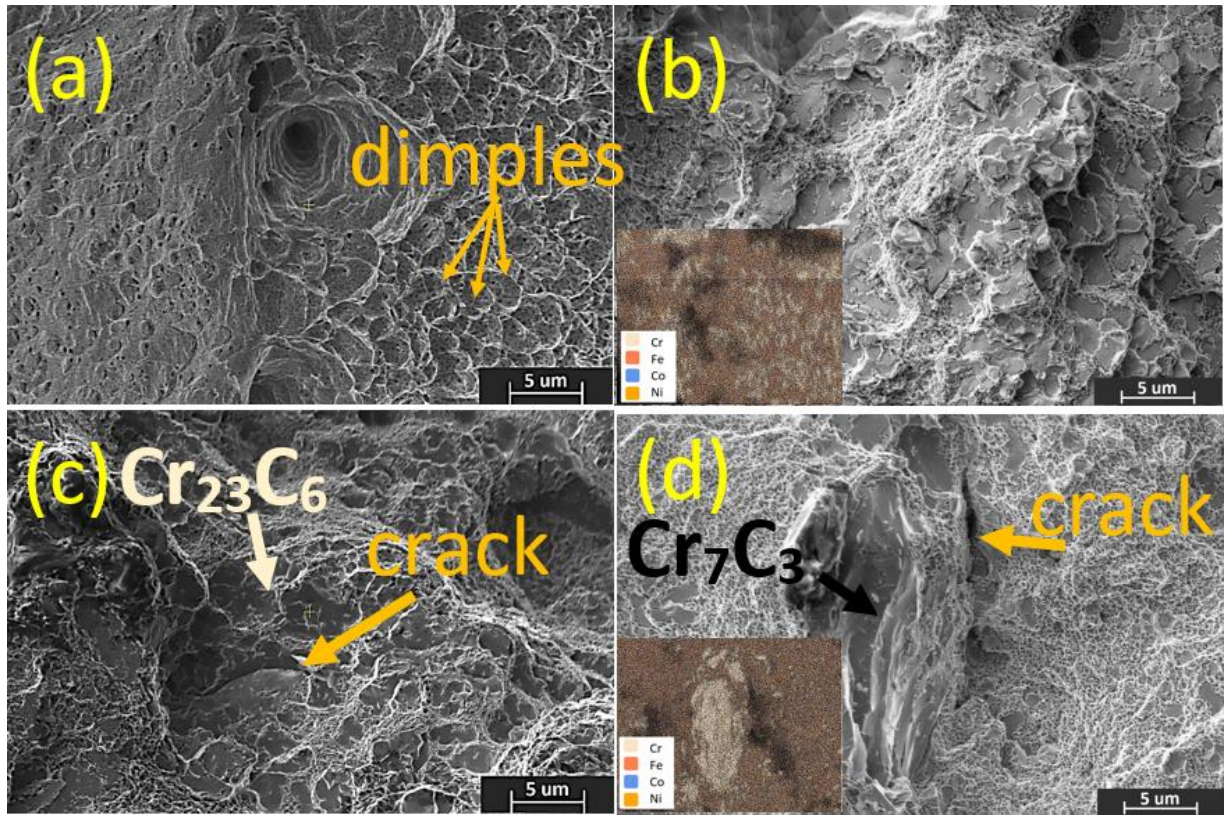

Fig. 8 SEM fracture topography images of (a) AM; (b) GA with an accumulated EDS map; (c) MA-850; (d) MA-1050 with an accumulated EDS map after room-temperature tensile tests.

For a clear comparison of deformation mechanisms at room temperature, EBSD maps are depicted in Fig. 9. The elongation of the grains in the deformation direction occurs in the AM and GA samples, as shown in Fig. 9 (a) and (b), respectively. Moreover, both MA samples (Fig. 9 (c-d)) present fine circular grains surrounded by carbide particles visible here as black spots. In turn, kernel average misorientation (KAM) diagrams demonstrate high dislocation agglomeration in all the samples. Similar dislocation densities are observed in both the AM (Fig. 9 (e)) and GA (Fig. 9 (f)) samples. Dislocation accumulation was detected at the grain and twin boundaries. However, a significantly lower number of dislocations mostly agglomerated in the grain boundaries, and secondary phases were visible in the MA samples (Fig. 9 (g-h)).

The EBSD maps of all the samples after tensile testing at 575°C are shown in Fig. 10. Notably, elliptical-shaped grains were detected in the AM samples (Fig. 10 (a)) with the minimal number of dislocations in the KAM diagrams (Fig. 10 (e)). On the other hand, grain elongation can be observed in the GA samples (Fig. 10 (b)), together with many deformed twins followed by dislocation agglomeration near the grain and twin boundaries. This suggests the presence of several additional strengthening mechanisms in the GA samples compared with those in the AM samples. A slight difference was also observed between MA-850 and MA-1050. In MA-850 (Fig. 10 (c)), a much lower number of twins was observed than in MA-1050 (Fig. 10 (d)). Additionally, grain elongation in the deformation direction was not detected. Moreover, the KAM maps demonstrated that dislocations accumulated at the grain boundaries and around the precipitates (Fig. 10 (g-h)).

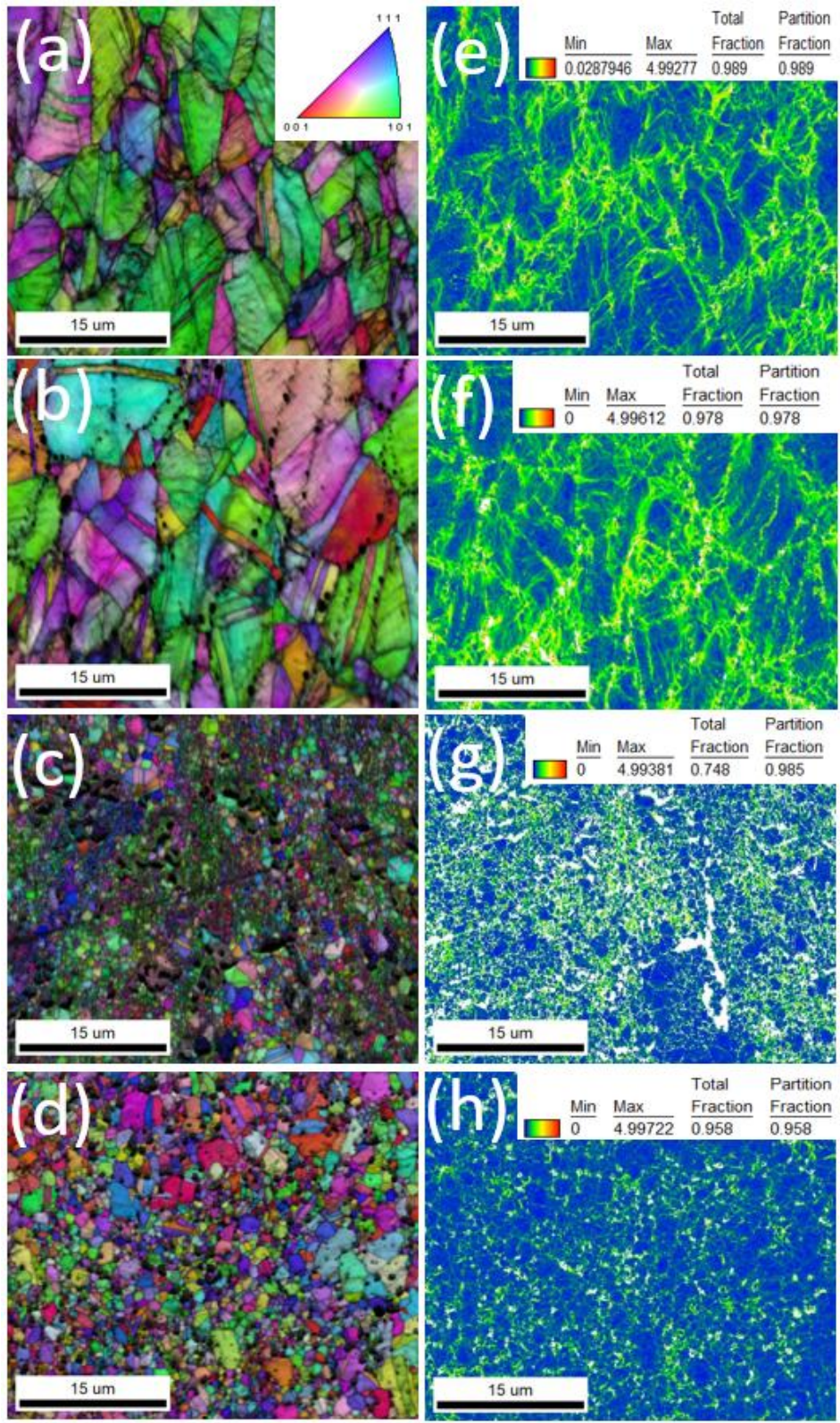

Fig. 9 Representative EBSD maps of all fractured samples at room temperature: (a) AM, (b) GA, (c) MA-850 and (d) MA-1050 with corresponding KAM maps: (e) AM, (f) GA, (g) MA-850 and (h) MA-1050.

To elucidate the underlying mechanisms at elevated temperatures, in-depth GA characterization via TEM was employed. The images are shown in Fig. 11. Both the grain boundaries and oxide precipitates act as barriers for dislocation movement, which greatly improves the mechanical properties (Fig. 11 (a)). The diffraction patterns in Fig. 11 (b) confirm the presence of the FCC phase as a matrix. The results above suggest a complex strengthening mechanism in the GA sample which is responsible for improving the mechanical properties at elevated temperatures compared with those of the AM sample, which is deprived of chromium oxides.

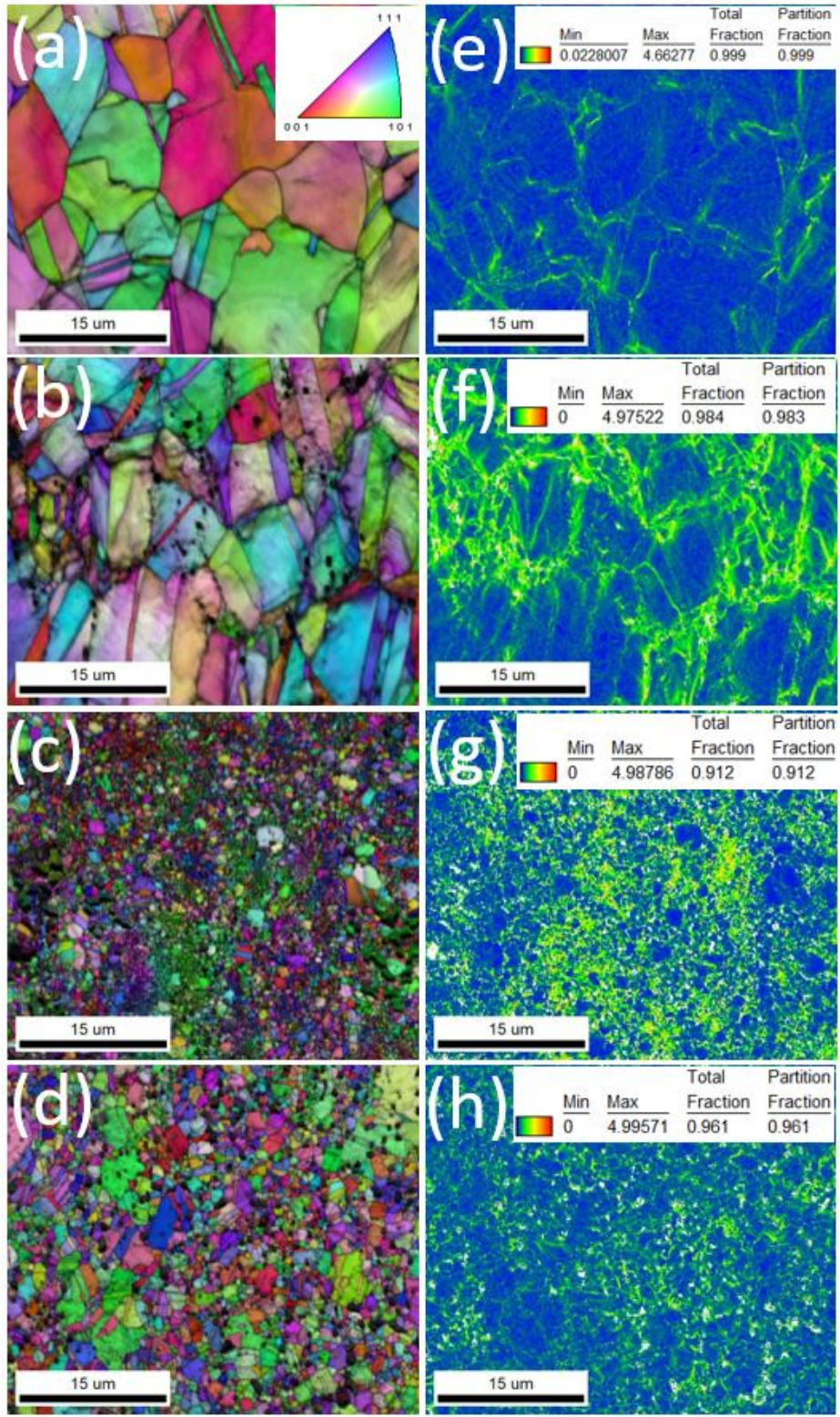

Fig. 10 Representative EBSD maps of all fractured samples at 575°C: (a) AM, (b) GA, (c) MA-850, (d) MA-1050 with corresponding KAM maps: (e) AM, (f) GA, (g) MA-850 and (h) MA-1050.

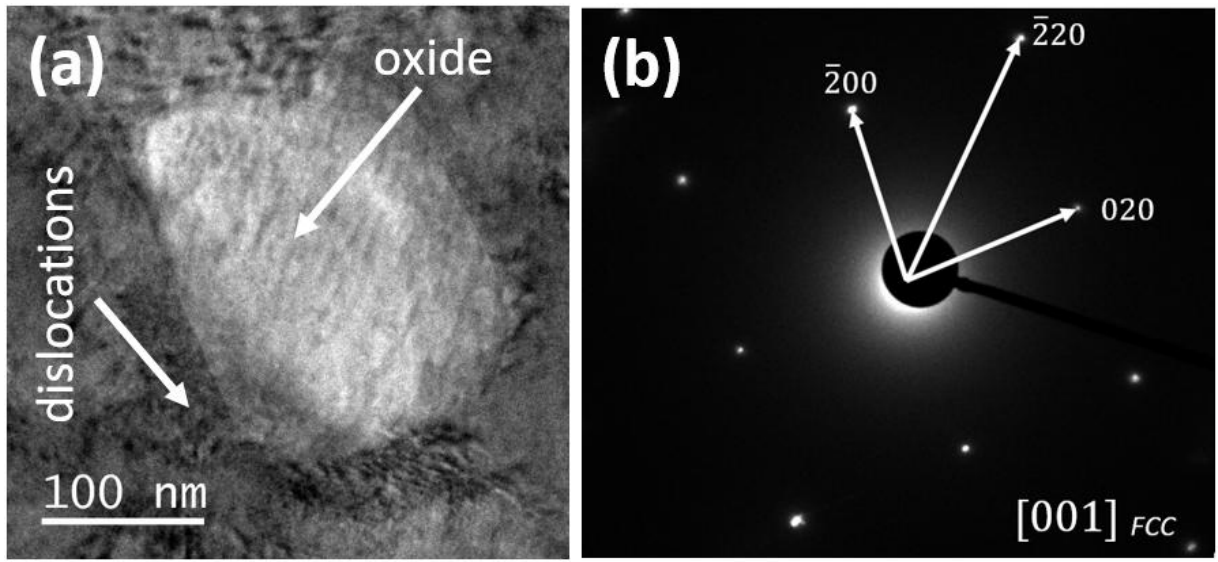

Fig. 11 (a) TEM images of the GA sample after tensile testing at 575°C and (b) SAED pattern of the FCC phase.

## 4. Discussion

This work shows that the manufacturing process significantly impacts the microstructure of CoCrNiFe MPEAs, which results in different mechanical properties related to the strengthening mechanisms caused by the presence of precipitates and grain size refinement. To prove these results, three different processing routes were chosen: arc melting followed by thermomechanical treatment, gas atomization + spark plasma sintering + annealing and mechanical alloying followed by spark plasma sintering and annealing. Moreover, two different annealing temperatures were used for the mechanically alloyed samples.

First, X-ray diffraction patterns were analyzed to obtain the phase compositions of all the manufactured samples, as demonstrated in Fig. 1. Different types of carbides were formed depending on the annealing temperature via the MA + SPS + annealing route: $Cr_{23}C_6$ at 850°C and $Cr_7C_3$ at 1050°C. Moreover, chromium oxides were formed in both samples. The AM and GA samples did not exhibit any strengthening phases. However, microstructural studies of the GA sample (see Figs. 2 and 5) revealed the presence of chromium oxides. As shown in the Supplemental file, the amount of oxide precipitates was below the XRD detection level. The full list of phase compositions for all the samples obtained via XRD and SEM/TEM is presented in Table 1. SEM studies confirmed single FCC phase formation in the AM sample. On the other hand, two types of carbides visible as light gray and dark gray precipitates in the MA samples were found. A comparison of the EBSD data from GA and AM, shown in Fig. 3, demonstrated that significantly finer grains were formed in the MA samples. The average grain sizes of all the samples are shown in Table 2. The EDS studies presented in Fig. 4 clearly show that the formation of secondary phases occurs at the expense of chromium in the matrix phase.

Two different tensile test conditions were chosen to study the mechanical properties of all the samples. Tests were performed at room temperature and 575°C, at which point carbide evolution was not expected. First, the room temperature tensile test results are shown in Fig. 6. A greater number of hard strengthening phases and a smaller grain size clearly improved the yield strength and ultimate tensile strength, but the elongation of the material decreased. In turn, the analysis at elevated temperatures can be found in Fig. 7. As expected, a higher temperature results in a decrease in the mechanical properties. The fractographs for all samples at room temperature in Fig. 8 confirm the ductile behavior of GA and AM and the brittle behavior of both MA samples. The EBSD data, shown in Fig. 9, associated with the KAM diagrams confirm that the dislocations are consumed mainly by the grain boundaries and precipitates.

EBSD images and KAM maps obtained after tensile tests at 575°C, as shown in Fig. 10, reveal a significantly lower number of dislocations in the AM sample. The MA samples did not differ substantially, whereas the GA samples consisted of many dislocations that were suppressed by grain boundaries, twin boundaries and oxide precipitates, as shown in the TEM image in Fig. 11.

In a future study, the thermal properties and irradiation resistance of all the samples will be tested. Based on these papers, the complex characterization and description of the impact of chromium-rich secondary phases on properties will be revealed.

### 4.1 Limitations of manufacturing processes

In further discussion, the AM sample has been treated as a benchmark due to the absence of secondary phases. In general, the implementation of casting/melting methods leads to the formation of dendritic structures. Consequently, other post-processes are needed to design the expected properties. In this work, homogenization, cold rolling, and recrystallization were chosen to obtain the desired grain structure, decrease the grain size, and limit the internal stress from plastic deformation. In the end, materials with fine grain structures were obtained. Furthermore, no secondary phase was found, which clearly demonstrated that the high-temperature heat treatment in air did not degrade the FCC phase.

In turn, in the implementation of the GA manufacturing route, slight microstructural evolution was observed. The main difference is the formation of chromium oxide rings in the grain boundaries or groups of grains. The shape of the rings might suggest that the oxide ring is in fact an oxide layer on each ex-particle. As mentioned before, a milling process was added before SPS to reduce the circular shape of the particles. The circular shape hinders air evacuation during SPS and leaves some air molecules between particles. The residual oxygen then reacts with Cr (the most oxidizing element among those present) and forms a thin layer between particles. The milling process decreased the circular shape, but the effect was not eliminated.

The use of the MA process results in the formation of carbides. Our previous work [44] clearly demonstrated the evolution of carbides after SPS and after annealing using the same annealing temperatures. At this point, carbide formation is difficult to avoid for technical reasons. As mentioned before, a process control agent is an organic solvent that is added to the milling jar to avoid cold welding. This, in turn, explains why PCA was not added to the GA sample. Process control agents prevent cold welding of powder particles and enable the alloying process. In the case of the GA, the powder particles were already alloyed, and cold welding was not prohibited. A reduction in the spheroidal shape might take place via an increase in the particle size. On the other hand, high temperatures inside the milling chamber caused the disintegration of all the organic solvents and the migration of C atoms. Carbon reacts with other alloying elements, especially Cr, which leads to the formation of chromium carbides. Previous works also revealed $Cr_{23}C_6$ -> $Cr_7C_3$ transformation at approximately 600°C [50,51]. Additionally, the use of WC milling media increases the possibility of wear-induced contamination and reactions, such as with Co to form composite carbides, e.g., $Co_3W_3C$. However, no such phases were detected via PXRD or EDS/TEM analyses in the current work.

Based on the Cr-C phase diagram [55] shown in Fig. S5 in the supplemental file, both $Cr_{23}C_6$ and $Cr_7C_3$ can exist in parallel. This phenomenon was also confirmed in our previous work [44]. However, the Cr-C binary system (attached in the Supplemental file) does not fully represent the case described in this paper. The evolution of carbides may largely depend on the local environment. Multiple elements are taken into these samples, which creates an additional difficulty. Moreover, previous work revealed a transformation from $Cr_{23}C_6$ to $Cr_7C_3$ at approximately 600°C [50,51]. This confirms that chromium carbides behave differently when they are distributed in a matrix phase. One can see that the heat treatment time and quenching cause the phenomena to be obscured in the diagram.

One can conclude that the reinforcing phases have great potential for improving material properties. They must be separated into two groups: *in situ*, which forms during the manufacturing process from the alloying elements and contaminations, and *ex situ*, which are added intentionally to the material to obtain the expected properties. The presence of secondary phases must always be considered, especially for powder metallurgy techniques.

**4.2 Effect of chromium oxides**

Owing to the relatively coarse grains and lack of precipitation strengthening, it can be assumed that chemical complexity is correlated with the strength of the AM sample [56]. This assumption is even strengthened because a solid solution strengthening effect is common in MPEAs, which often favors them over conventional alloys [1–3]. In this context, a slight evolution of the microstructure associated with the presence of chromium oxides in the GA sample with a constant grain size does not lead to remarkable mechanical properties at RT. The increase in UTS is related only to an extra barrier for dislocation motion.

Tensile tests conducted at elevated temperatures revealed decreased mechanical properties for both the AM and GA samples. Wu et al. confirmed that the decrease in the YS at elevated temperatures for MPEAs may be determined by the lattice friction dominated by Peierls barriers [57]. A comparison of the experimental results with the Peierls–Nabarro calculations confirms that the barriers in compositionally complex alloys are greater than those in conventional alloys. Moreover, when examining the fracture area, EBSD and KAM maps clearly reveal that dynamic recrystallization occurs [24]. Consequently, the energy delivered into the material during the tensile test was at least partially consumed by the reorganization of the crystal lattice. In turn, a limited number of dislocations were found. This observation strictly corresponds to the limited deformation ability and strengthening effect, as observed by Wu [57].

A comparison of the results described above with those of the GA sample clearly revealed that even a slight introduction of fine oxide precipitates greatly improved both the strength and elongation at elevated temperatures. First and foremost, TEM images and KAM maps reveal that fine precipitates act as an extra barrier for dislocation motion (Fig. 11 (a)), which is typical for the precipitation-strengthening effect [58,59]. However, based on the analysis performed at room temperature, this effect is not crucial because of the limited amount of the secondary phase.

On the other hand, the EBSD maps associated with the KAM images clearly indicate that the chromium precipitates greatly suppressed the dynamic recrystallization process, as shown in Fig. 10 (b), (f). The grains visible in the EBSD maps remained elongated in the deformation direction. Furthermore, more dislocations can be found in the KAM maps, not only near the precipitates but also near the grain boundaries. This means that more energy might be stored inside the material before fracture, and as a consequence, the ultimate tensile strength and elongation increase.

In the investigation of the deformation mechanism occurring in the studied samples, the shape of the stress-strain curves should be considered. As mentioned in the previous section, different types of serrations occurred in the AM and GA samples [54,57,60]. The AM sample shows constantly occurring 'impulses,' suggesting dynamic strain aging (DSA) of type A (sudden increases in flow stress followed by sharp dips, owing to shear band nucleation and Lüders band movement, with origins in interstitial atom agglomeration around dislocations as they move and break free from these Cottrell atmospheres) [54,60], as shown in Fig. 7 (b). At this moment, the serration behavior observed in the stress-strain curves should be considered. Owing to the chemical complexity of MPEAs, the most popular models of serration description should be rejected, resulting in the consideration of different models. To explain the recorded data, we compared them with existing results in the literature. In the case of the Cottrel model [54,61], implementation of this model was impossible because of the trace amounts of interstitial atoms in the MPEAs. In addition, the whole concept of an interstitial or otherwise smaller atom atmosphere (Cottrell or Cottrell-like) causing DSA is invalid when there is no one dominant base element and a low interstitial concentration, as atmospheres of smaller MPEA atomic constituents are unlikely to naturally form akin to Type A DSAs. Compared to the kinetics of dislocation motion, their diffusion is often too slow to quickly form Cottrell-like atmospheres, unlike that of carbon in low-alloy steels. On the other hand, in the Sleeswyk model [54,61], atomic diffusion through a moving dislocation is hindered by the very well-known sluggish diffusion effect. As a consequence, Tsai et al. proposed a different model in which the reorganization of elements near the dislocation occurs [61]. This effect, while perhaps Cottrell-like in nature, proceeds more slowly owing to the aforementioned sluggish diffusion and could be considered its own mechanism, which is unique to MPEAs with high concentrations of not necessarily fast moving, yet smaller atomic constituents. This, in turn, limits the local energy and pins the dislocations. This means that when a dislocation is blocked, more energy is required to restart its movement. In this way the effect manifests like that of Type A DSAs, but its underlying mechanism is fundamentally different owing to higher concentrations of slower moving, not necessarily interstitial, MPEA constituents. This energy accumulation is visible as the stress increases in the stress-strain curve, as shown in Fig. 7 (b), which is marked with an arrow. After the dislocation is released, the dislocation is more mobile, which is visible as a temporary stress drop. As the stress increases, more defects are generated. Consequently, more time is needed to overcome these obstacles. Furthermore, the dynamic recrystallization process is visible in the EBSD map in Fig. 10 (a). This leads to more chaotic serrations in the stress-strain curve in Fig. 7 (c) than in the more regular and predictable serrations found in Type A DSAs. Finally, the GA sample demonstrates different serration effects, similar to D-type serration behavior [54], with multiple plateaus, or staircases, shaped features in the stress strain curve without Type A-like decreases in stress. However, D-type serrations generally occur at relatively low temperatures and/or high strain rates. This might suggest a different underlying mechanism to explain the features in the stress/strain curves of the GA sample.

As expected, conducting tensile tests at high temperatures resulted in a decrease in each of the mechanical parameters. The decreases in yield strength at 575°C for the GA and AM samples are clearly similar (~30%). On the other hand, the UTS and elongation, which decreased for the AM sample (35% and 53%, respectively), are significantly greater than those for the GA sample (18% and 14%, respectively). Based on this, it can be assumed that an additional strengthening mechanism occurs in the AM sample, which is related to the UTS and elongation improvement. In contrast, the increase in the YS is related mostly to the effects described above: recrystallization suppression and precipitation strengthening.

An explanation for the abovementioned doubts might be found in the EBSD maps in Fig. 10 (b). As mentioned previously, a significantly greater number of twins in the GA samples than in the AM samples were revealed. This suggests the presence of a twinning-induced plasticity (TWIP) effect in oxide-reinforced CoCrFeNi MPEAs [62,63]. The TWIP effect confirms the strain hardening rate curve for the GA samples, where several peaks demonstrating substantial improvement in hardening were found. These peaks may provide strong evidence for twinning formation at elevated temperatures.

**4.3 TWIP effect**

The TWIP mechanism is still not fully understood, especially for MPEAs. However, some core information has already been described. First, Choi et al. demonstrated that the elastoplastic transition is affected only by dislocation slip [64]. Moreover, Kalidindi et al. reported that for TWIP steel, the initial stage of the strain hardening rate curve is fully related to dislocation slip [65]. In this stage, the twinning effect does not occur, which was also confirmed by Asgari et al. [66]. However, some stacking faults can be found, which might be nuclei for further twin boundaries. The fact that the initial stage of deformation is fully controlled by dislocation slip is beyond the scope of this discussion. This effect explains the similar percentage drop in yield strength of AM and GA; the TWIP mechanism does not control the elastic region.

The problem appears in the next stages. Kalidindi et al. reported that the activation of dislocation slip and mutual overlap lead to the formation of twin boundaries, resulting in ~3–4% strain [65]. On the other hand, Gutierrez-Urruria and Raabe postulated that the main strengthening factor in the next stage is the strong accumulation of dislocations and the formation of dislocation cells or walls [63,67]. In conclusion, the impact of twin boundaries should be investigated in the following stage.

One can say that dislocation slip is the main strengthening effect throughout the deformation process. Twinning is only an additional factor but might be key for improving mechanical properties [62,63]. In this context, many researchers have reported that a low stacking fault energy is a great parameter for enhancing the TWIP effect [26,63]. Therefore,

CoCrFeNi, where the stacking fault energy is 20 mJ/m$^2$, seems to be one of the best candidates for realizing this mechanism in practice [26].

Based on the assumptions presented above, an explanation of the serration behavior of the GA sample at elevated temperatures is highly important. Recent papers have shown that similar Portevin-Le Chateliers (PLCs) might be the result of the TWIP effect [41,42]. Lebedkina et al. proposed that the steps in stress-strain curves in the DSA region are related to twinning [70]. These authors suggested that deformation twins cause internal stresses and local hardening, which drive the persistent propagation of localized deformation bands. Consequently, a similar behavior to that shown in Fig. 7 (d) occurs. Notably, the PLC effect in the AM sample was found at ~4% elongation, whereas serration in the GA sample was observed after an elongation of 12%. This finding indicates that strong defect accumulation results in serration. Notably, oxide precipitates do not promote twinning but can affect the TWIP effect by hindering the recrystallization process.

**4.4 Effect of chromium carbides**

The combination of a large plastic region with a satisfactory tensile strength is a great point for further microstructural development, as it represents the often sought, yet rarely realized, realization of toughness by simultaneous increases in strength and ductility. These two properties are frequently in tension with each other, and breaking the tradeoff is the central theme of broader impact in this study and many others. In this work, the impacts of two types of chromium carbides were analyzed. The discussion presented above points to the conclusion that $Cr_7C_3$ is a significantly better option in terms of improving the mechanical properties. The analysis of the stress-strain curves at room temperature revealed that the UTS of CoCrFeNi reinforced with $Cr_{23}C_6$ was only slightly greater than that of $Cr_7C_3$. Considering that MA-850 has a twofold smaller average grain size, which should be an advantage for UTS, the hypothesis of a positive effect of $Cr_7C_3$ is further strengthened, as shown in Fig. 6. Moreover, MA-850 breaks immediately after reaching the YS, whereas MA-1050 presents a small, but visibly recorded plastic region.

The positive impact of $Cr_7C_3$ is even more visible when tensile tests are conducted at elevated temperatures. The YS, UTS, and elongation decrease in the case of MA-1050 are 14%, 17% and 9%, respectively, whereas for MA-850, the decreases are significantly greater. There are at least two potential reasons for these observations. First, our recent work revealed that $Cr_7C_3$ has a hardness three times lower than that of $Cr_{23}C_6$ [44]. This, in turn, might promote slight plastic deformation. On the other hand, tensile tests at 575°C may facilitate some thermal effects. In addition, in MA-1050, some twin boundaries formed in the $Cr_7C_3$-reinforced CoCrFeNi MPEA. This suggests that twinning also occurs in MA-1050, which is promising for further optimization of the amount of chromium carbides.

The average grain sizes of MA-850 and MA-1050 are significantly smaller than those of GA and AM. This effect might promote an increase in the YS and UTS, together with a decrease in the impact of carbides. However, the formation of chromium carbides occurred at the expense of the Cr content in the matrix phase, where chromium visibly improved the mechanical properties. Many publications have shown that the addition of chromium to FCC-MPEAs results in significant strength/hardness improvement [6,25,71]. For example, Yan et al [6] demonstrated that the modification of the Cr content from 0 to 21 wt.% caused the UTS to increase at RT from 486 MPa to 559 MPa for the $Al_{0.3}Cr_xFeNiCo$ MPEA. Moreover, the elongation even increases from 41 to 47%, indicating another success at breaking the tension between the strength and ductility trade-offs. The abovementioned results show that more chromium might be added to the milling jar to promote equiatomic CoCrFeNi alloys with some chromium carbide precipitates to account for Cr loss from the matrix. In this way, processing this MPEA can be thought of as akin to steel processing, where one must add excess Cr beyond the desired composition to account for evaporation and loss during melting and ingot fabrication. However, this might take place separately in each case when the manufacturing process is optimized. Notably, the carbide clusters, not the finely distributed carbides, were the weakest points of the MA samples.

Bearing the abovementioned factors in mind, it might be assumed that the amount of chromium carbide and carbide clusters are the greatest obstacles to MA sample improvement. The process optimization must be performed in two different fields. First, limiting the PCA, or the usage of different PCAs, might be beneficial for increasing the Cr/Cr-carbide ratio. Furthermore, the implementation of longer pre-milling or exact milling processes should promote even chromium carbide distribution. This, in turn, results in improved mechanical properties with satisfactory elongation maintenance. Notably, this process optimization might lead to plastic region broadening, causing some extra effects on the stress-strain curves. Elevated temperatures might promote the migration of carbon atoms through grain boundaries and form another barrier for dislocation movement [72], restoring the Cottrell atmospheres and leading to Type A DSAs. As a consequence, different types of DSA effects than those in the GA might occur.

As presented in this work, different reinforcing phases were formed depending on the manufacturing process. These phases are difficult to avoid because of their *in situ* formation. However, the presence of these phases is not necessarily a disadvantage, and one could use them to improve the mechanical properties, as shown in some of the cases investigated in this study. The presented work provides a detailed description of their impact on properties and provides insights for further composition modification and understanding potential challenges in the process. Moreover, in this work we demonstrated that *in situ*-formed reinforcing phases may constitute a great alternative for improving materials. However, this can occur only if the process is fully under control.

**4.5 Scalability challenges**

The scalability of all processes and their potential industrial application for MPEAs should be considered. As mentioned above, arc melting requires a high vacuum, which at least partially limits the scalability of this process. On the other hand, high corrosion resistance has been confirmed in the literature [5]. Consequently, we decided to perform the heat treatment processes in a traditional muffle furnace with only Ar rinsing. Indeed, microstructural and diffraction images exclude the potential formation of secondary phases (especially chromium oxides). This in turn points to the conclusion that traditional casting or melting techniques often used in industry might be implemented. Notably, these observations concern the Co-Cr-Fe-Ni system. Other elements, especially highly oxidized ones, must be checked separately. The manufactured component will require not only thermo-mechanical treatment but also some machining processes.

In turn, the gas atomization technique is a significantly less productive process than melting or casting. However, the powder may be produced in a continuous process, which significantly eases process optimization [73]. Furthermore, a thin oxide layer might form on the surface of powder particles, especially at high temperatures, as demonstrated in this work. On the other hand, as proposed in this work, fine oxide particles may have a positive effect on the mechanical properties at room and elevated temperatures.

The formation of powder via mechanical alloying poses the greatest risk of potential secondary phase formation, as clearly demonstrated in this work. On the other hand, after sintering, the samples present significantly finer grains. This might improve the tensile strength if the process is fully controlled. In this case, the capacity of milling jars is limited; however, it is still growing.

In the end, the sintering via SPS used for the three samples in this work is interesting because of the short processing time, which hinders secondary phase formation [39]. Additionally, SPS provides an opportunity for the shape formation of a manufactured product [1]. Powder compaction enables the intentional addition of secondary phases, a considerable advantage over melting and casting techniques. Consequently, the number of powder metallurgy applications continues to grow, confirming that it is a promising direction for modern manufacturing techniques, including MPEAs.

## 5. Conclusion

In summary, the impacts of different manufacturing processes and different strengthening mechanisms on the mechanical properties of CoCrFeNi MPEAs were analyzed. Two different tensile test conditions were chosen, i.e., room temperature and 575°C.

1. A slight introduction of chromium oxides through the FCC matrix caused a small difference in the YS and UTS at RT but a significant difference at elevated temperatures. This occurred because an additional effect of dynamic recrystallization suppression and a TWIP effect are associated with the strengthening of the precipitates.

2. Different types of serrations were observed depending on the presence or absence of chromium oxides. It was confirmed that twins act as additional obstacles for dislocation movement, which reflects different DSA effects.

3. The presence of chromium carbides drastically decreases the elongation with increasing strength. However, chromium carbide clusters were found to be the weakest point of these composite structures. The strengthening effect is more pronounced at room temperature.

4. $Cr_7C_3$ was found to be a more promising strengthening phase than $Cr_{23}C_6$. This is related to the significantly lower hardness of this material.

5. The introduction of chromium oxides and chromium carbides might be a promising way to improve the mechanical properties when the manufacturing process is fully controlled. The large elongation region of the gas-atomized sample provides an opportunity for the addition of carbides and strength improvement. On the other hand, the presence of chromium oxides promotes the TWIP effect, which has an additional effect at elevated temperatures.

**Declaration of competing interest**

The authors declare that they have no known competing financial interests or personal relationships that could have appeared to influence the work reported in this paper.

**Data Availability**

Data will be made available on request.

**Acknowledgements**

We acknowledge support from the European Union Horizon 2020 research and innovation program under NOMATEN Teaming grant (agreement no. 857470) and from the European Regional Development Fund via the Foundation for Polish Science International Research Agenda PLUS program grant No. MAB PLUS/2018/8. Ł.K. acknowledges support from Polish National Agency for Academic Exchange, which supported this research via the Bekker NAWA Programme. The publication was created within the framework of the project of the Minister of Science and Higher Education 'Support for the activities of Centres of Excellence established in Poland under Horizon 2020' under contract no. MEiN/2023/DIR/3795.